\documentclass[journal=ancac3,manuscript=article,layout=twocolumn]{achemso}
\usepackage{multirow}
\usepackage{chemformula} 
\usepackage[T1]{fontenc} 
\usepackage{hyperref}
\usepackage{comment}
\usepackage[version=4]{mhchem}
\usepackage{soul}
\usepackage{lettrine}
\usepackage{siunitx}
\usepackage{amssymb}

\author{Djardiel da S. Gomes}
 \affiliation{State University of Campinas, Gleb Wataghin Institute of Physics, Department of Applied Physics, 13083-859, Campinas, São Paulo, Brazil.}
 \alsoaffiliation{University of Bras\'{i}lia, Faculty UnB Planaltina, Materials Science Postgraduate Program, Bras\'{i}lia, Federal District, Brazil.}

\author{Isaac M. Felix}
 \affiliation{Center for Agri-food Science and Technology, Federal University of Campina Grande, 58840-000, Pombal, Paraíba, Brazil.}

\author{Lucas L. Lage}
 \affiliation{Fluminense Federal University, Institute of Physics, Av. Litorânea, 24210-340, Niterói, Rio de Janeiro, Brazil.}
 
\author{\\Douglas S. Galvão}
\affiliation{Group of Organic Solids and New Materials (GSONM), Gleb Wataghin Institute of Physics, University of Campinas (UNICAMP), Campinas, SP, Brazil.}
\alsoaffiliation{Center for Computational Engineering \& Sciences (CCES), University of Campinas - UNICAMP, Campinas, SP, Brazil.}

\author{Andrea Latgé}
 \affiliation{Fluminense Federal University, Institute of Physics, Av. Litorânea, 24210-340, Niterói, Rio de Janeiro, Brazil.}

\author{Marcelo L. Pereira Junior}
 \affiliation{University of Bras\'{i}lia, College of Technology, Department of Electrical Engineering, 70910-900, Bras\'{i}lia, Federal District, Brazil.}
 \alsoaffiliation{University of Bras\'{i}lia, Faculty UnB Planaltina, Materials Science Postgraduate Program, Bras\'{i}lia, Federal District, Brazil.}

 \email{marcelo.lopes@unb.br}

\title[]
  {
Topology as a Design Variable for Multiproperty Engineering in Synthesized 4-5-6-8 Carbon Nanoribbons
  }


\begin{document}

\begin{abstract}
Nonbenzenoid carbon frameworks expand the design landscape of low-dimensional materials by introducing controlled departures from hexagonal symmetry. Here, we demonstrate that the experimentally realized 4-5-6-8 carbon nanoribbon establishes a topology-driven paradigm for multiproperty engineering rather than representing a simple structural variant of graphene nanoribbons. Combining hybrid density functional theory, parametrized tight-binding, and molecular dynamics within a coherent multiscale framework, we show that the symmetry-broken lattice stabilizes a hierarchy of bonds while remaining in the same energy range with ribbons of comparable width. This geometric organization produces a robust semiconducting state with a hybrid-functional  electronic band gap exceeding 1 eV and enables strain to function as a controllable parameter for electronic modulation. Notably, a tight-binding Hamiltonian fitted only at equilibrium, accurately captures the strain-dependent electronic band evolution, indicating that the essential physics is dominated by the topology itself. Mechanical analysis reveals high stiffness with fracture governed by the largest polygonal motifs, demonstrating that geometric asymmetry redistributes stress without compromising structural integrity. In addition, intrinsic phonon scattering suppresses lattice thermal conductance, allowing favorable thermoelectric performance to emerge without extrinsic disorder. The optical response further confirms that nonequivalent ring connectivity reorganizes interband transitions, promoting strong absorption within the visible range and efficient photocarrier generation. These results position the  topology as a governing physical parameter capable of coupling elasticity, electronic structure, thermal transport, and optical activity, establishing the 4-5-6-8 nanoribbon as a unified platform for the predictive design of multifunctional carbon materials.
\end{abstract}

\vspace{0.5cm}

\section{Introduction}

\lettrine[lines=2, findent=0pt, nindent=2pt]{T}{}opology has emerged as a powerful organizing principle to control the physical behavior of low-dimensional materials \cite{xiao2007valley,ren2016topological,culcer2020transport}. In carbon-based systems, where bonding versatility enables a rich diversity of allotropes, altering the polygonal structure of the lattice provides a direct pathway for tailoring electronic states beyond the constraints imposed by hexagonal symmetry \cite{xiao2007valley,liu2017graphene}. Moving away from the benzenoid paradigm is therefore not merely a structural modification, but a strategy capable of unlocking fundamentally new regimes of quantum behavior \cite{lombardi2019quantum,chen2024topological}.

Graphene nanoribbons (GNRs) represent one of the most successful platforms for bandgap engineering through dimensional confinement \cite{son2006energy,karakachian2020one,gomes2025computational}. Their electronic properties are governed by atomic-scale geometry, including width \cite{karakachian2020one,merino2017width}, edge termination \cite{hod2007enhanced,chang2014geometric,zhang2013experimentally,akhmerov2008boundary}, and boundary conditions \cite{akhmerov2008boundary,brey2006electronic,yazyev2013guide}, enabling tunable semiconducting behavior suitable for nanoelectronic applications \cite{bennett2013bottom,celis2016graphene,krishnan2023graphene}. It should be stressed that, although certain topological analogies can be drawn with graphene nanowiggles (GNWs) \cite{cai2010atomically,bizao2017mechanical}, such systems constitute a distinct class of nanostructures with different structural origins. Yet, despite these advances, most GNR designs remain rooted in hexagonal networks, where topology acts primarily as a secondary descriptor rather than a deliberate design variable.

A fundamentally different paradigm arises when hexagonal symmetry is intentionally disrupted by the incorporation of nonhexagonal rings \cite{zhang2013structures,fan2021biphenylene,tong2022ultrahigh,qu2025surface}. Theoretical studies have long suggested that carbon lattices containing tetragonal, pentagonal, and octagonal motifs can host a wide spectrum of electronic responses, spanning metallic, semimetallic, and semiconducting phases \cite{tong2022ultrahigh,wang2015phagraphene,mortazavi2023theoretical,ma2025non}. However, the experimental realization of atomically precise architectures embedding multiple polygonal units has historically posed a major challenge.

This limitation was recently overcome with the on-surface synthesis of a carbon nanoribbon composed of 4-5-6-8-membered rings via lateral fusion of polyfluorene chains on Au(111) \cite{kang2023surface}. High-resolution scanning probe microscopy confirmed its atomic structure and revealed a semiconducting bandgap of approximately 1.4 eV, establishing nonbenzenoid topology as an experimentally accessible design space \cite{kang2023surface}. Despite this breakthrough, a central question remains unanswered: can such topological complexity serve as a predictive parameter to engineer coupled physical responses rather than merely modifying isolated properties?

Here, we address this question by demonstrating that the deliberate rupture of hexagonal symmetry acts as a topological mechanism for emergent multiphysical behavior in one-dimensional carbon systems. Rather than treating the 4-5-6-8 nanoribbon as an isolated structural realization, we establish the nonbenzenoid topology as a unifying design parameter that simultaneously governs electronic structure, quantum transport, and thermoelectric performance.

To uncover this physics, we have adopted a multiscale computational architecture that bridges atomistic-scale topology with transport-relevant length scales. Hybrid-functional density functional theory (DFT) is employed to obtain an accurate electronic description consistent with experimental observations. These results are subsequently mapped onto a parametrized tight-binding Hamiltonian, enabling large-scale quantum transport calculations while preserving the essential electronic features. The lattice thermal conductivity is further evaluated using reverse nonequilibrium molecular dynamics, allowing charge and heat transport to be assessed within a unified theoretical framework.

Our results reveal that geometric frustration, bond heterogeneity, and symmetry breaking collectively generate strongly coupled electronic and thermal responses, positioning topology not as a passive structural attribute but as an active topological-tune control for functional materials design. By positing nonbenzenoid architectures from synthetic achievements to predictive platforms, this work establishes a conceptual foundation for the rational engineering of next-generation carbon nanostructures.

\section{Results and Discussion}

\begin{figure*}[t!]
\centering
\includegraphics[width=0.9\linewidth]{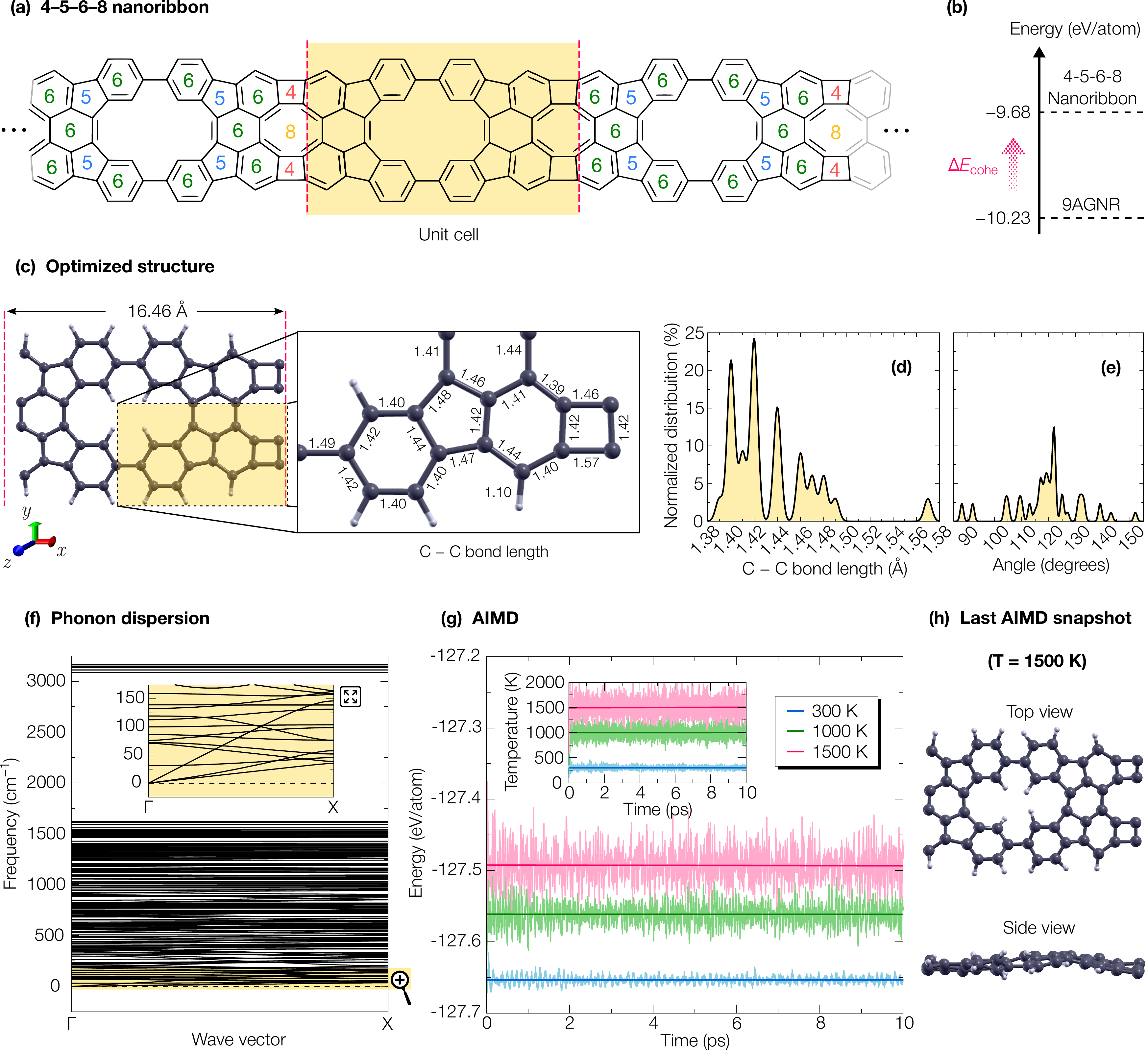}
\caption{
Structural characterization and stability of the 4-5-6-8 carbon nanoribbon. 
(a) Atomic structure highlighting the periodic incorporation of nonhexagonal rings within the ribbon backbone. 
(b) Cohesive energy comparison with a width-matched armchair graphene nanoribbon, indicating similar energy stability. 
(c) Optimized geometry and representative C-C bond lengths, revealing pronounced bonding heterogeneity induced by the mixed polygonal network. 
(d,e) Statistical distributions of bond lengths and bond angles, thereby providing a consistent framework for analyzing a uniform hexagonal environment. 
(f) Phonon dispersion confirming dynamical structural stability through the absence of imaginary modes. 
(g) \textit{Ab initio} molecular dynamics simulations at multiple temperatures demonstrating structural thermal robustness. 
(h) Final ab initio molecular dynamics snapshot at 1500 K, showing preservation of the ribbon morphology.
}
\label{fig:fig1}
\end{figure*}

We begin by examining the structural consequences imposed by the nonbenzenoid topology, since the lattice's geometrical topology ultimately constrains the stability window within which electronic and transport phenomena emerge. Figure~\ref{fig:fig1} provides a multiscale structural description of the 4-5-6-8 nanoribbons, connecting atomic tiling, bonding statistics, lattice dynamics, and finite-temperature behavior. Rather than representing a perturbed/distorted graphene ribbon, this system defines a topologically distinct carbon phase in which tetragons and octagons are periodically embedded into a predominantly $sp^2$ network.

The backbone shown in Fig.~\ref{fig:fig1}(a) reproduces the structural motif resolved experimentally through on-surface synthesis, where scanning probe measurements confirmed the periodic arrangement of 4-, 5-, 6-, and 8-membered rings. This synthetic route demonstrates that the ribbon is not a metastable theoretical construct but an experimentally accessible topology stabilized through controlled cyclodehydrogenation reactions. Establishing whether such a symmetry-broken lattice remains energetically competitive with graphene-derived nanoribbons is therefore an important and necessary analysis.

To quantify the relative structural stability, we evaluated the cohesive energy per atom, defined as:

\begin{equation}
E_{\mathrm{coh}} =
\frac{E_{\mathrm{tot}} - \sum_i n_i E_i^{\mathrm{atom}}}{N},
\end{equation}

where $E_{\mathrm{tot}}$ is the total energy of the relaxed unit cell, $E_i^{\mathrm{atom}}$ is the energy of an isolated atom of species $i$, $n_i$ is its multiplicity, and $N$ is the total number of atoms. Figure~\ref{fig:fig1}(b) compares the cohesive energy of the 4-5-6-8 nanoribbon with that of a 9AGNR of similar width~\cite{Fujita1996}, which serves as a meaningful benchmark given the strong width dependence of quasi-one-dimensional carbon systems. The results indicate that the nonbenzenoid ribbon is slightly less stable than its hexagonal counterpart, reflecting the energetic cost associated with angular frustration introduced by tetragons and octagons. Importantly, the magnitude of this difference remains small, confirming that the intentional breakdown of hexagonal symmetry does not structurally destabilize the lattice to a prohibitive extent but instead preserves energetic viability within the same confinement regime that stabilizes conventional armchair graphene nanoribbons.

This interpretation aligns well with first-principles simulation results reported by Mortazavi \cite{mortazavi2023theoretical}, who identified bond heterogeneity as an intrinsic structural feature of pristine 4-5-6-8 nanoribbons. In that work, nitrogen substitution at the ribbon edges produced a measurable contraction of the lattice and redistributed the mechanically critical C-C bonds surrounding the nonbenzenoid junctions, demonstrating that the edge's chemistry directly couples to the global strain landscape. The agreement between these trends and current structural metrics supports the conclusion that the observed bonding hierarchy arises from topology rather than computational artifacts.

The optimized structure shown in Fig.~\ref{fig:fig1}(c) clarifies how the lattice changes balance the competing angular constraints imposed by mixed polygons. The unit-cell length of approximately 16.46~\AA\ defines the longitudinal periodicity, while the highlighted region reveals a broad distribution of C-C bond lengths ranging from about 1.38 to 1.58~\AA. In contrast to graphene and narrow AGNRs, which typically exhibit nearly uniform bonding, the present topology redistributes strain across multiple ring types, generating localized regions of compression and extension that collectively stabilize the network.

The statistical distributions in Fig.~\ref{fig:fig1}(d) and Fig.~\ref{fig:fig1}(e) provide a fingerprint of these structural accommodations. The bond-length histogram exhibits multiple maxima rather than the single dominant peak characteristic of hexagonal carbon, confirming that structural heterogeneity is topologically enforced. The angular distribution is equally revealing. The absence of a pronounced peak near $90^\circ$ indicates that the four-membered rings do not adopt ideal rectangular geometries but instead undergo angular distortion that partially relieves strain through coupling with adjacent pentagon and octagon rings. Similar distortions and the identification of specific load-bearing bonds have been reported before in the literature \cite{mortazavi2023theoretical}, further supporting the view that the mechanical resilience in these ribbons arises from cooperative bond rearrangements around the nonbenzenoid cores.

The comparison with the 9AGNR highlights the structural implications of symmetry breaking. Whereas the honeycomb lattice tightly constrains geometric degrees of freedom in armchair ribbons, the 4-5-6-8 topology introduces an internal strain landscape while preserving the dominant $sp^2$ character, as evidenced by the persistence of bond angle distributions around $120^\circ$. The resulting backbone is therefore structurally heterogeneous yet still electronically sp2-type conjugated, a combination difficult to achieve through conventional graphene patterning.

The dynamical structural stability is confirmed by the phonon dispersion presented in Fig.~\ref{fig:fig1}(f), where no imaginary frequencies are present throughout the entire Brillouin zone, indicating that the ribbon resides at a true minimum of the potential-energy surface. The highest-frequency optical modes extend beyond approximately 3000~cm$^{-1}$ and are primarily associated with C-H stretching vibrations localized at the ribbon edges, while intermediate-frequency modes originate from in-plane C-C stretching within the backbone. The absence of soft modes indicates that the geometric frustration introduced by the nonbenzenoid topology does not lead to vibrational instability.

Thermal robustness was further assessed through ab initio molecular dynamics. As shown in Fig.~\ref{fig:fig1}(g), the total energy remains stable (within the limit of the thermal fluctuations) at temperatures up to 1500~K. They fluctuate around stable average values without signatures of bond break or large-scale structural reconstructions. The final configuration displayed in Fig.~\ref{fig:fig1}(h) confirms the preservation of the backbone, with only slightly out-of-plane structural corrugations, which are typical of atomically thin materials. Comparable thermal stability of both pristine and chemically modified variants has been reported \cite{mortazavi2023theoretical}, supporting the interpretation that the mixed-ring topology defines a mechanically robust carbon phase once formed.

From an experimental standpoint, this stability is particularly relevant because the synthesis protocol involves annealing steps that activate polymer fusion and cyclodehydrogenation. The ability of the ribbon to withstand substantially higher simulated temperatures suggests that the nonbenzenoid backbone constitutes a resilient metastable configuration compatible with experimentally accessible growth conditions.

Taken together, Fig.~\ref{fig:fig1} demonstrates that the 4-5-6-8 topology simultaneously accommodates geometric distortion, energy viability, and dynamical structural stability. Breaking hexagonal symmetry does not compromise the lattice integrity. Instead, it generates a hierarchy of bonds and angles that stabilizes the structure while introducing internal degrees of freedom absent in conventional graphene nanoribbons.

\begin{figure*}[t!]
    \centering \includegraphics[width=.9\linewidth]{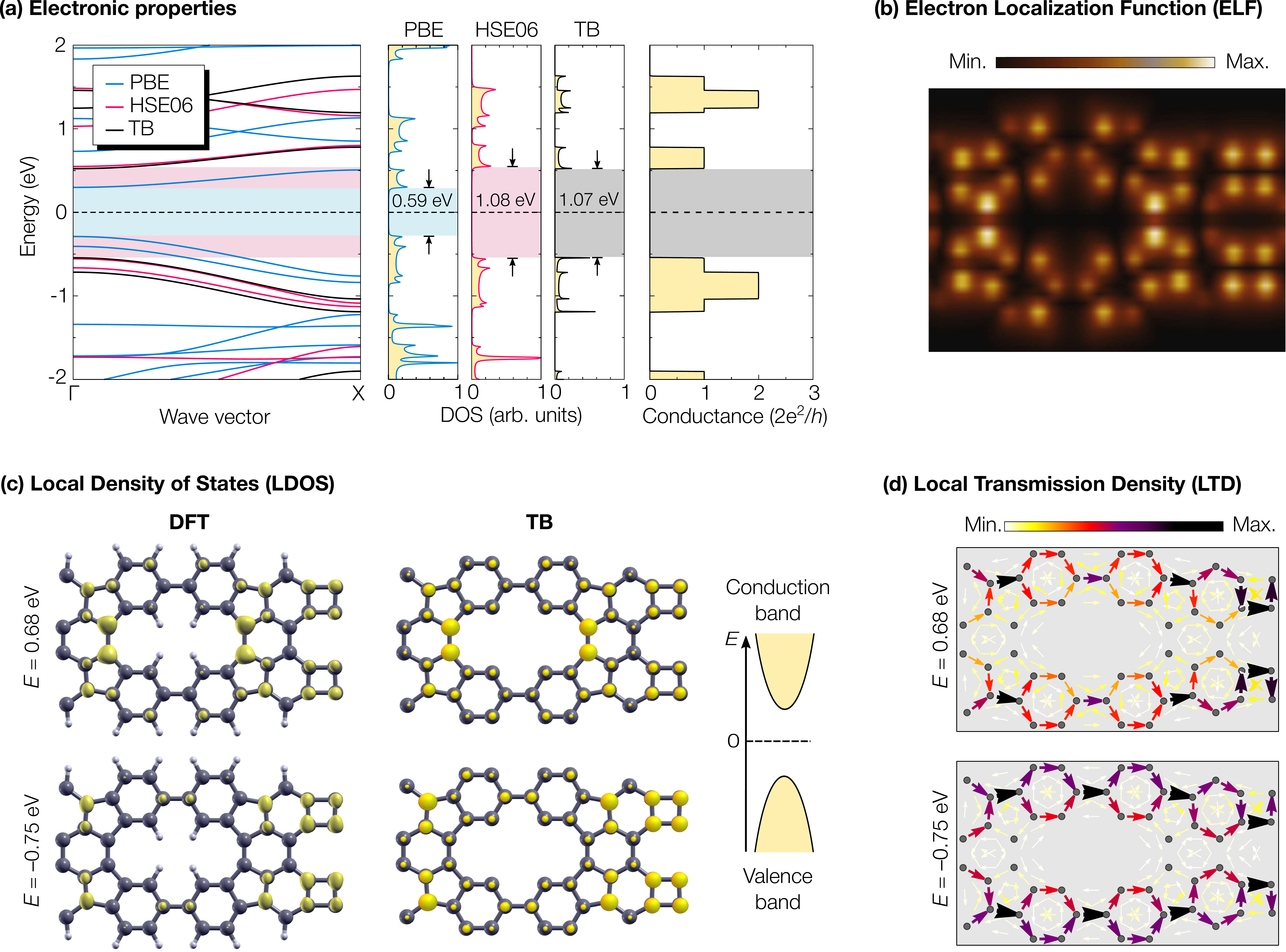}
    \caption{Electronic and transport properties. (a) Electronic structure overview showing the band structure (left), the corresponding Density of States (DOS) (middle), and quantum conductance (right). The results are compared among the PBE (blue), HSE06 (red), and Tight-Binding (TB, black) methods. The calculated electronic band gaps are 0.59 eV (PBE), 1.08 eV (HSE06), and 1.07 eV (TB), respectively. (b) Electron Localization Function (ELF) map. (c) Local Density of States (LDOS) comparing DFT and TB spatial distributions. (d) Local Transmission Density (LTD) showing bond currents. Both (c) and (d) are visualized at the valence band peak ($E = -0.75$ eV) and conduction band valley ($E = 0.68$ eV). }\label{fig:fig2}
\end{figure*}

These experimentally realizable, thermally robust structural features provide the physical foundation for understanding the emergent electronic and transport properties discussed next. In conjugated carbon systems, the electronic structure is inseparable from the underlying bonding network; consequently, the strain redistribution and bond hierarchy imposed by the mixed-ring topology are expected to reshape the $\pi$-electron manifold directly. Figure~\ref{fig:fig2} examines this connection by combining hybrid-functional electronic structure, tight-binding reconstruction, and real-space transport analysis within a unified framework. Rather than behaving as a perturbed/distorted armchair graphene nanoribbon, the 4-5-6-8 lattice defines an electronically distinct regime in which symmetry breaking becomes an active parameter for electronic band engineering.

All methods reveal that the electronic band structures shown in Fig.~\ref{fig:fig2}(a) are of semiconductors, while simultaneously exposing the importance of exchange-correlation treatment. The PBE functional predicts an electronic band gap of approximately 0.59~eV, whereas the screened hybrid functional increases this value to about 1.08~eV, correcting the typical gap underestimation of semi-local approximations and placing the electronic scale closer to experimentally relevant energies. Hybrid-functional calculations previously reported for this topology exhibit the same qualitative trend, reinforcing that the gap opening is not a methodological artifact but a robust electronic feature of the nonbenzenoid backbone. The tight-binding Hamiltonian, parametrized directly from the HSE06 dispersion, reproduces the band gap value of the hybrid functional (1.07~eV) while preserving the curvature of the frontier bands. This hierarchical consistency demonstrates that the multiscale strategy is not a redundant stacking of techniques, but a physically necessary pathway that transfers hybrid-functional accuracy to transport-accessible length scales.

From the perspective of graphene nanoribbon physics, the magnitude and origin of this gap are particularly instructive. A 9AGNR of comparable width typically belongs to the $3p$ family, where the band gap emerges primarily from quantum confinement within an otherwise hexagonal lattice \cite{Fujita1996}. The 4-5-6-8 ribbon cannot be mapped onto any of the canonical armchair families ($3p$, $3p+1$, $3p+2$), because the periodic insertion of nonhexagonal rings breaks the translational symmetry that underpins that classification. The different geometry leads to a wider band gap, since orbital overlap reorganization is driven by angular frustration. Thus, the lattice topology serves as a tool to promote electronic localization and reduce transport channels compared with 9AGNR near the Fermi level \cite{Fujita1996}.


The density of states in Fig.~\ref{fig:fig2}(a) further highlights this departure from conventional armchair behavior. Instead of the symmetric van Hove singularities characteristic of graphene nanoribbons, the spectrum exhibits uneven spectral weight near the band edges, indicating that the valence and conduction states originate from nonequivalent bonding environments. This asymmetry is a direct electronic manifestation of the bond heterogeneity established in Fig.~\ref{fig:fig1}, confirming that structural frustration propagates into the low-energy electronic landscape.

Insight into the orbital organization is provided by the electron localization function shown in Fig.~\ref{fig:fig2}(b). Regions of strong localization trace the shorter C-C bonds surrounding pentagon-hexagon junctions, whereas comparatively depleted charge density appears near the more weakly bonded segments associated with the larger rings. The resulting pattern indicates that $\pi$ electrons remain globally conjugated but are spatially modulated by the internal strain field. Such redistribution establishes preferential pathways for carrier motion while preserving electronic continuity along the backbone.

Hybrid-functional studies of chemically modified variants of this ribbon have previously shown that nitrogen substitution alters both the electronic band gap and charge distribution by introducing electron-rich centers that locally reshape the frontier states without significantly disrupting conjugation. The fact that comparable electronic anisotropy already emerges in the pristine structure indicates that topology alone places the system near a regime of electronically tunable frustration, with heteroatom incorporation acting as a secondary refinement rather than a prerequisite for band engineering.

Real-space projections of the frontier states, presented in Fig.~\ref{fig:fig2}(c), confirm that both band edges are dominated by $p_z$ orbitals forming an extended $\pi$ network across the ribbon interior. Enhanced amplitudes are consistently observed near the nonbenzenoid junctions, identifying these regions as electronic hot spots where curvature in the bonding topology spatially concentrates electronic wavefunctions. This behavior establishes a meaningful parallel with experimental scanning-probe observations, which revealed that states near the band edges propagate along the ribbon backbone instead of being localized exclusively at the edges, consistent with a bulk-like conduction channel embedded within a quasi-one-dimensional geometry.

The transport consequences of this orbital structure are captured by the conductance profile in Fig.~\ref{fig:fig2}(a). Quantized steps emerge as discrete propagating modes become available, and the rapid increase in conductance indicates that multiple channels are opened within a narrow energy window. In contrast to a 9AGNR, whose transmission is largely dictated by width-controlled subbands, the present ribbon benefits from internally generated conduction pathways arising from topological heterogeneity. The mixed-ring lattice, therefore, behaves as a self-patterned electronic medium in which geometry preconfigures transport.

This interpretation is reinforced by the local density of states and transmission maps. The LDOS distributions in Fig.~\ref{fig:fig2}(c) reveal that frontier states remain spatially continuous, avoiding the strong edge confinement that can limit mobility in narrow graphene nanoribbons. Meanwhile, the local transmission density shown in Fig.~\ref{fig:fig2}(d) traces percolating current pathways that adapt to the polygonal arrangement, bending around pentagon and octagon rings while maintaining global directionality. Carrier flow is therefore redirected rather than suppressed by the broken symmetry.

The electronic response thus emerges directly from the symmetry-broken bonding network introduced by the 4-5-6-8 topology. By redistributing orbital overlap across nonequivalent polygons, the lattice reshapes the density of states and promotes spatially extended transport channels that are inaccessible within purely hexagonal ribbons. In this framework, topology ceases to be a geometric descriptor and instead becomes a governing physical parameter that couples band formation to carrier propagation, establishing a robust platform for multiproperty engineering in low-dimensional carbon systems.


\begin{figure*}[t!]
    \centering \includegraphics[width=.9\linewidth]{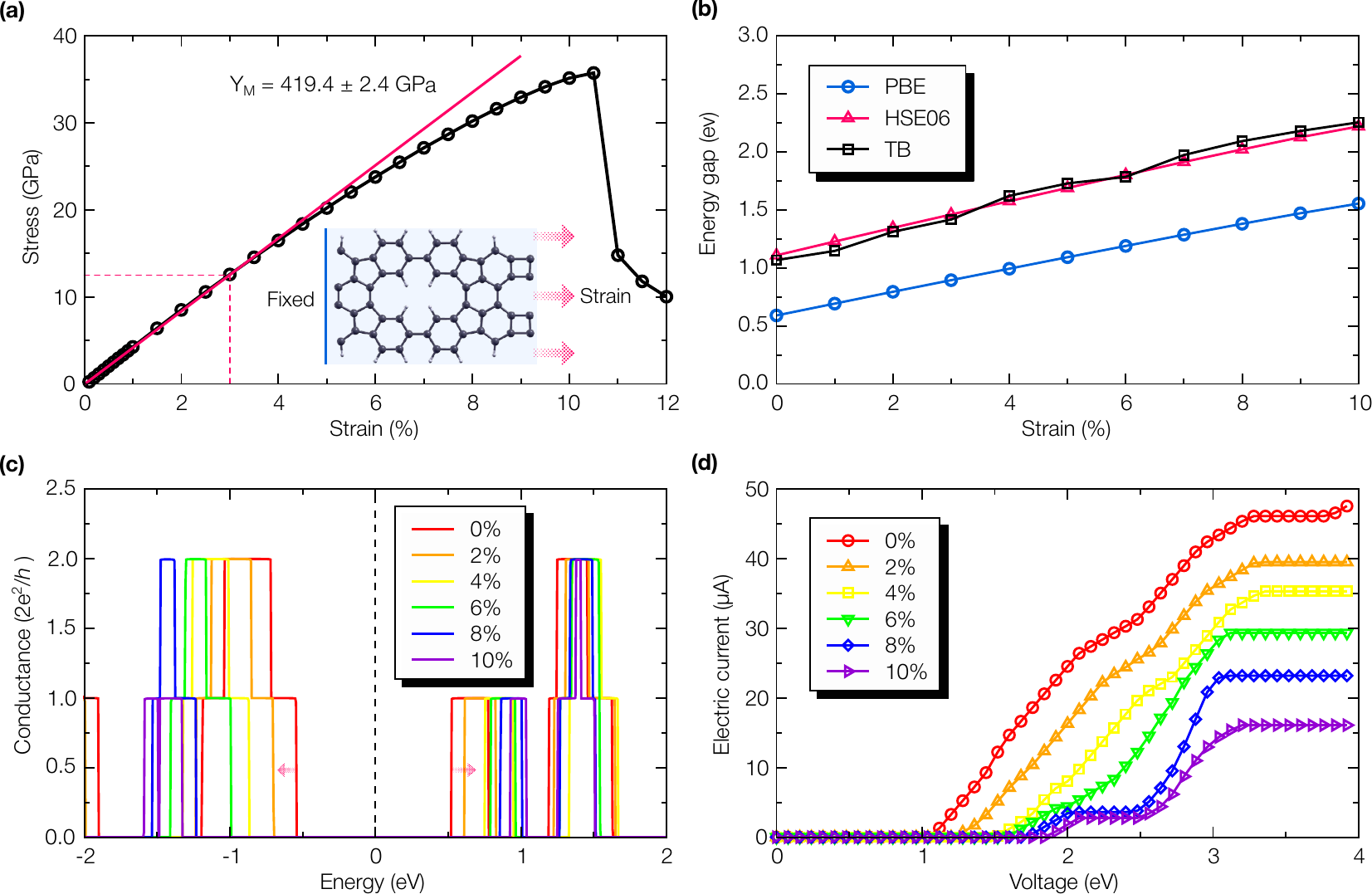}
    \caption{Selected physical properties of 4-5-6-8 nanoribbon as a function of an externally applied uniaxial strain at $T=0$ K. (a) DFT-based stress-strain curve showing an estimated Young’s modulus of $Y_{M} = 419.4 \pm 2.4$ GPa. (b) Evolution of the energy gap versus strain, comparing ab initio calculations (PBE in blue, HSE06 in red) with Tight-Binding (TB in black) results. (c) Electronic conductance and (d) electric current as a function of bias voltage, calculated using the TB method for strain strengths ranging from 0\% to 10\%.}
    \label{fig:fig3}
\end{figure*}

Figure~\ref{fig:fig3}(a) shows that the 4-5-6-8 nanoribbon exhibits a well-defined linear elastic regime, from which a Young’s modulus of $419.4 \pm 2.4$ GPa is estimated. Although smaller than the stiffness typically reported for pristine graphene, this magnitude firmly places the ribbon within the class of mechanically resilient carbon nanostructures. The reduced modulus should be inferred as a direct consequence of bond heterogeneity rather than structural fragility. While hexagonal lattices distribute stress nearly uniformly, the mixed-ring topology introduces bonds with distinct stiffnesses, enabling the structure to redistribute elastic energy through multiple deformation pathways, but less effectively than the hexagonal topology.

The fracture mechanism provides a particularly clear demonstration of how topology determines the mechanical response. Structural failure nucleates at the largest interior ring, precisely where the density of load-bearing bonds is lowest and angular distortion is highest. First-principles simulations have shown that two symmetric C-C bonds located at the center of this nanopore dominate the tensile strength, effectively defining the mechanical limit of the ribbon fracture. Nitrogen substitution near these bonds reduces the tensile strength from approximately 40 GPa to approximately 33 GPa, confirming that local electronic perturbations directly weaken the primary stress-transfer channel while leaving the global framework intact. Thermal simulations further indicate that both pristine and nitrogen-terminated ribbons remain structurally stable at elevated temperatures, demonstrating that the predicted failure mode reflects intrinsic topology rather than thermally activated degradation.

This behavior contrasts with the fracture patterns typically observed in armchair graphene nanoribbons, where rupture frequently begins at edges or defect sites. In the current system, failure is encoded in the internal geometry itself. Symmetry breaking, therefore, does not introduce mechanical weakness; instead, it spatially programs the locus of rupture, transforming topology into a predictive descriptor of mechanical robustness.

Once mechanical integrity is established, strain can be viewed not as a destructive perturbation but as a controllable thermodynamic variable that can reshape the electronic structure. Figure~\ref{fig:fig3}(b) reveals a monotonic widening of the band gap with increasing tensile strain for all methods. The hybrid functional shifts the electronic scale upward relative to the PBE, while the tight-binding Hamiltonian, parametrized only once from the unstrained hybrid dispersion, reproduces the strain dependence with remarkable accuracy. This agreement indicates that the governing physics is encoded in the connectivity of the nonhexagonal backbone rather than in functional-specific details, allowing hybrid-level fidelity to propagate into transport-relevant regimes without the need of iterative refitting.

In standard graphene nanoribbons, the electronic band gap modulation is largely interpreted through quantum confinement, with the band gap scaling inversely with ribbon width. The 4-5-6-8 lattice exhibits a distinct regime. Because the frontier states are already influenced by angular frustration and bond hierarchy, strain acts on a prestructured orbital network. Topology and mechanical deformation, therefore, act cooperatively, establishing a dual control mechanism for electronic band engineering, not available in purely hexagonal systems.

The transport response confirms this interpretation. Quantized conductance plateaus in Fig.~\ref{fig:fig3}(c), given by 1  and 2 units of quantum conductance, shift systematically with strain, tracking the displacement of the band edges, while the current-voltage characteristics in Fig.~\ref{fig:fig3}(d) exhibit progressively lower currents as the electronic band gap widens. Tensile loading thus provides a continuous means of regulating carrier injection without disrupting coherent transport channels. The resulting changes are reflected in the electric current, which requires an additional 1 eV at $10\%$ strain to activate the response compared to the unstrained case.
The decreasing current values as the strain increases from $0\%$ to $10\%$, resulting from a reduction in available transport states, are clearly shown in the current values. The saturation current at higher voltages drops from $45\mu A$ to $15\mu A$. 

Mechanical deformation, electronic structure, and carrier transport consequently emerge as interdependent manifestations of the same symmetry-broken bonding architecture. The 4-5-6-8 topology cannot merely withstand strain; it converts it into a predictable electronic response while preserving structural resilience. Topology, therefore, advances beyond a geometric motif to serve as a functional design parameter that couples elasticity to electronic band engineering in low-dimensional carbon materials.

\begin{figure*}[t!]
    \centering \includegraphics[width=.9\linewidth]{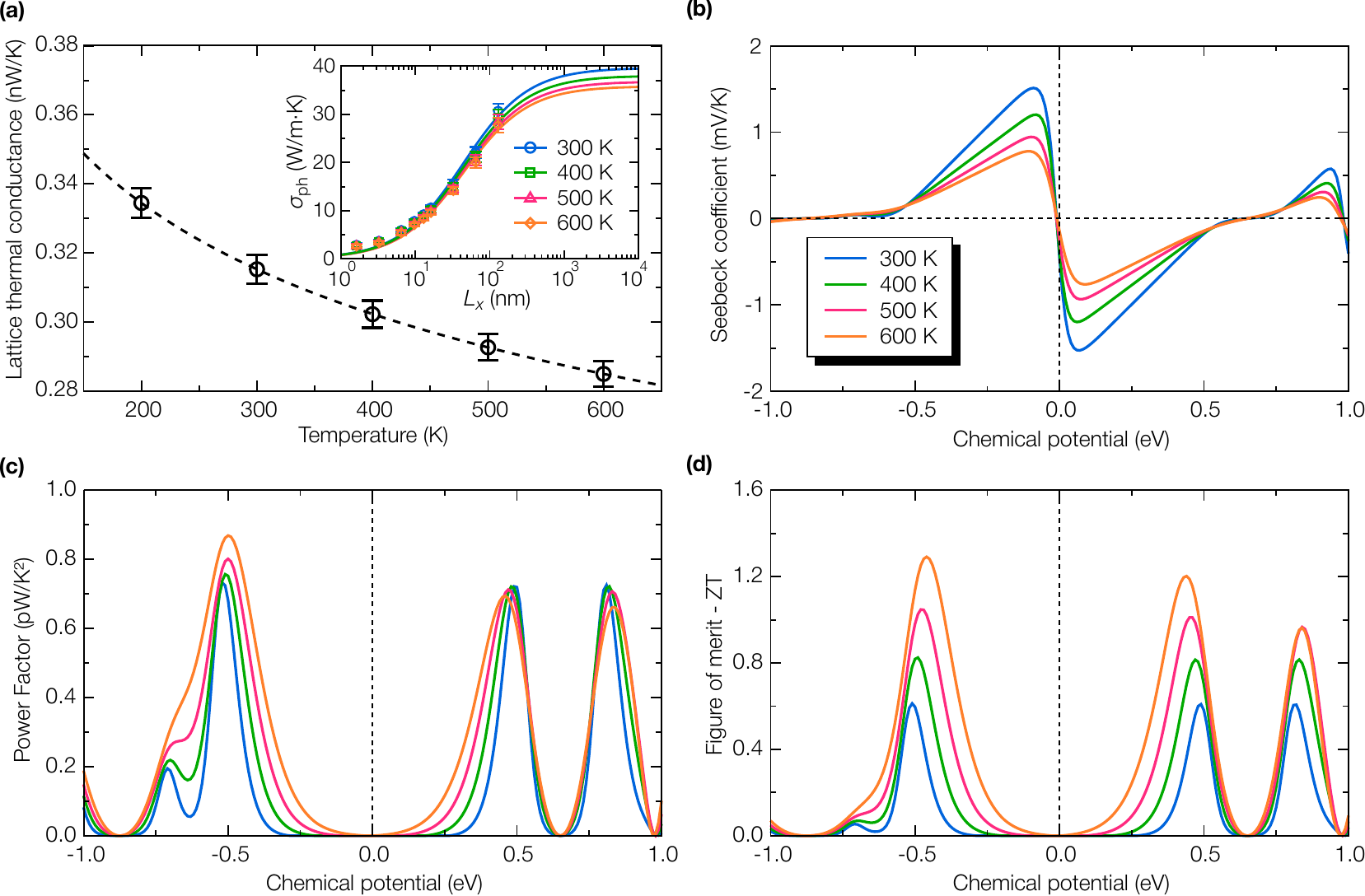}
    \caption{Thermal and thermoelectric transport properties. (a) Lattice thermal conductance ($\kappa_{ph}$) as a function of temperature, calculated using molecular dynamics. The inset shows the phonon thermal conductivity ($\sigma_{ph}$) as a function of length ($L$). (b) Seebeck coefficient, (c) Power factor, and (d) thermoelectric figure of merit ($ZT$) plotted as a function of chemical potential. The curves represent different temperatures ranging from $300$ K to $600$ K.}
    \label{fig:fig4}
\end{figure*}

If topology simultaneously governs electronic dispersion and mechanical adaptability, a natural question follows: Does the same structural principle also regulate energy flow? Thermoelectric performance provides a stringent test of this hypothesis because it depends on the delicate balance between carrier mobility and phonon-mediated heat transport. Fig.~\ref{fig:fig4} addresses this interplay by combining lattice thermal conductance obtained from molecular dynamics with tight-binding electronic transport, thereby extending the multiscale framework from atomistic stability to device-relevant thermodynamic regimes.

Fig. \ref{fig:fig4}(a) reveals a monotonic reduction of the ballistic lattice thermal conductance ($\kappa_\text{ph}^0$) with increasing temperature, a parameter derived from the length-dependent thermal conductivity scaling through Eq. \eqref{eq:conductance-conductivity}. This behavior underscores that the topological framework, which was previously shown to govern lattice stability and the electronic response to strain, serves as the fundamental regulator of heat transport. The observed temperature-dependent trend reflects the influence of bond heterogeneity on lattice dynamics, defining the upper bound for energy transport prior to the onset of diffusive scattering.

The inset of Fig. \ref{fig:fig4}(a) depicts the thermal conductivity of the 4-5-6-8-NR as a function of the characteristic ribbon length, $\sigma_{ph}(L_x)$, across various temperatures. By applying an extrapolation scheme for the infinite-length limit based on the ballistic-to-diffusive transition (Eq. \eqref{eq:length-dependence}), the intrinsic thermal conductivity $\sigma_\text{ph}^\infty$ was found to be $39.7 \pm 1.6$ W/mK at 300 K. This decrease is comparable to that observed in GNRs with pores \cite{gomes2025computational} isotopic doping \cite{jiang2010isotopic} chemisorption functionalization \cite{chien2012influence} and Stone-Thrower-Wales defects \cite{ng2012molecular}.
To provide a systematic quantitative benchmark and isolate the effect of hexagonal symmetry breaking, identical simulations were performed for armchair graphene nanoribbons (A-GNR) of equivalent width, which exhibited a significantly higher intrinsic conductivity of $231.2 \pm 9.3$ W/mK at 300 K, in agreement with previous reports \cite{chien2012influence,liu2018phonon,kim2021thermal}.

A direct comparison confirms that the periodic topological defects in the 4-5-6-8-NR lattice suppress thermal conductivity by more than 82\% relative to pristine A-GNRs. This pronounced decrease is attributed to enhanced phonon scattering at the pentagonal and octagonal rings, which act as intrinsic scattering centers.

Fig.~\ref{fig:fig4}(a) reveals a monotonic reduction of the lattice thermal conductance with increasing temperature, reflecting the progressive activation of phonon-phonon scattering channels. More revealing than the temperature trend itself is the magnitude of the thermal response. In pristine graphene, thermal conductivity can exceed several thousand W\,m$^{-1}$\,K$^{-1}$, a consequence of long phonon mean free paths and the absence of strong scattering centers \cite{balandin2008superior}. Such exceptional heat transport, while beneficial for thermal management, is fundamentally detrimental to thermoelectric efficiency because it suppresses the figure of merit \cite{snyder2008complex}.

The mixed-ring backbone alters this paradigm at the structural level. By redistributing bond lengths and introducing angular frustration across nonequivalent polygons, the 4-5-6-8 topology generates intrinsic phonon scattering without relying on extrinsic disorder, edge roughness, or isotope engineering. Heat transport is therefore attenuated as a direct consequence of lattice geometry. Rather than treating phonon suppression as a post-synthetic optimization strategy, the present architecture embeds it into the bonding network itself.

The inset of Fig.~\ref{fig:fig4}(a) further clarifies the transport regime. The length dependence of $\sigma_{ph}$ captures the expected crossover from ballistic to diffusive propagation, enabling extrapolation towards the intrinsic conductivity limit. This behavior closely parallels that of graphene nanoribbons, in which phonon mean free paths strongly influence size-dependent thermal transport, yet the absolute scale is reduced here by the lattice's structural complexity. The topology, therefore, operates as an internal regulator of heat flow, effectively shortening the phonon propagation length while preserving crystalline order.

Suppressing lattice thermal conductance alone does not guarantee favorable thermoelectric behavior. The decisive factor is whether electronic transport remains sufficiently robust. The Seebeck response shown in Fig.~\ref{fig:fig4}(b) provides strong evidence that this condition is satisfied. The pronounced antisymmetric profile around the chemical potential reflects the semiconducting character established earlier, while the increase in peak magnitude at lower temperatures follows the sharpening of the electronic distribution near the band edges.

The power factor displayed in Fig.~\ref{fig:fig4}(c) consolidates this picture. Its maxima occur near the band edges, where rapid variation of the transmission function boosts thermopower while maintaining appreciable electrical conductance. Importantly, the temperature evolution indicates that carrier transport remains stable across the thermal explored window, reinforcing that the electronic backbone is not compromised by the structural distortions that suppress phonons.

These combined effects culminate in the figure of merit shown in Fig.~\ref{fig:fig4}(d). The emergence of elevated $ZT$ values should not be interpreted as an isolated transport outcome, but as the macroscopic expression of a deeper organizing principle. Once hexagonal symmetry is relaxed, the lattice acquires internal degrees of freedom that mediate charge and heat transport along partially decoupled channels. Electronic states remain sufficiently delocalized to sustain transport, whereas phonon propagation is intrinsically disrupted by bond hierarchy and polygonal diversity. In fact, our findings contrast with the behavior of conventional two-dimensional graphene, whose gapless nature results in minimal thermopower. 

They significantly outperform armchair semiconducting nanoribbons of comparable width, such as the 9AGNR. 
In contrast, 9AGNRs exhibit a room-temperature ZT factor of approximately $0.25$~\cite{tran_optimizing_2017}, whereas the 4-5-6-8 nanoribbon reaches about $0.6$. This enhancement arises from a higher Seebeck coefficient at room temperature in the nanoribbon, attributable to their larger band gap compared to that of the 9AGNR.
Nanostructuring is typically required to open an electronic bandgap and enhance the Seebeck coefficient. Here, symmetry breaking accomplishes this electronically before any geometric confinement is invoked. The result is a system in which thermopower is amplified without sacrificing coherent transport pathways.

From a design perspective, this result marks an important conceptual shift. Conventional routes toward high thermoelectric efficiency in graphene-based materials typically rely on vacancy engineering, superlattices, isotope disorder, or hybrid edge structures to reduce thermal conductivity. In contrast, the present ribbon achieves the same objective solely through topology. The symmetry-broken lattice simultaneously reshapes the electronic spectrum, lacking electron-hole symmetry, preserves mechanical resilience, and suppresses lattice heat transport.

Thermoelectric behavior therefore emerges not as an auxiliary property layered onto a preexisting nanoribbon, but as a natural consequence of the same structural principle that governs bonding, elasticity, and electronic dispersion. The 4-5-6-8 topology transforms geometric asymmetry into a multiphysics control parameter, demonstrating that deliberate departures from hexagonal order can provide a unified pathway toward engineering coupled electronic and thermal functionalities in low-dimensional carbon systems.

\begin{figure*}[t!]
\begin{center}
\includegraphics[width=0.9\linewidth]{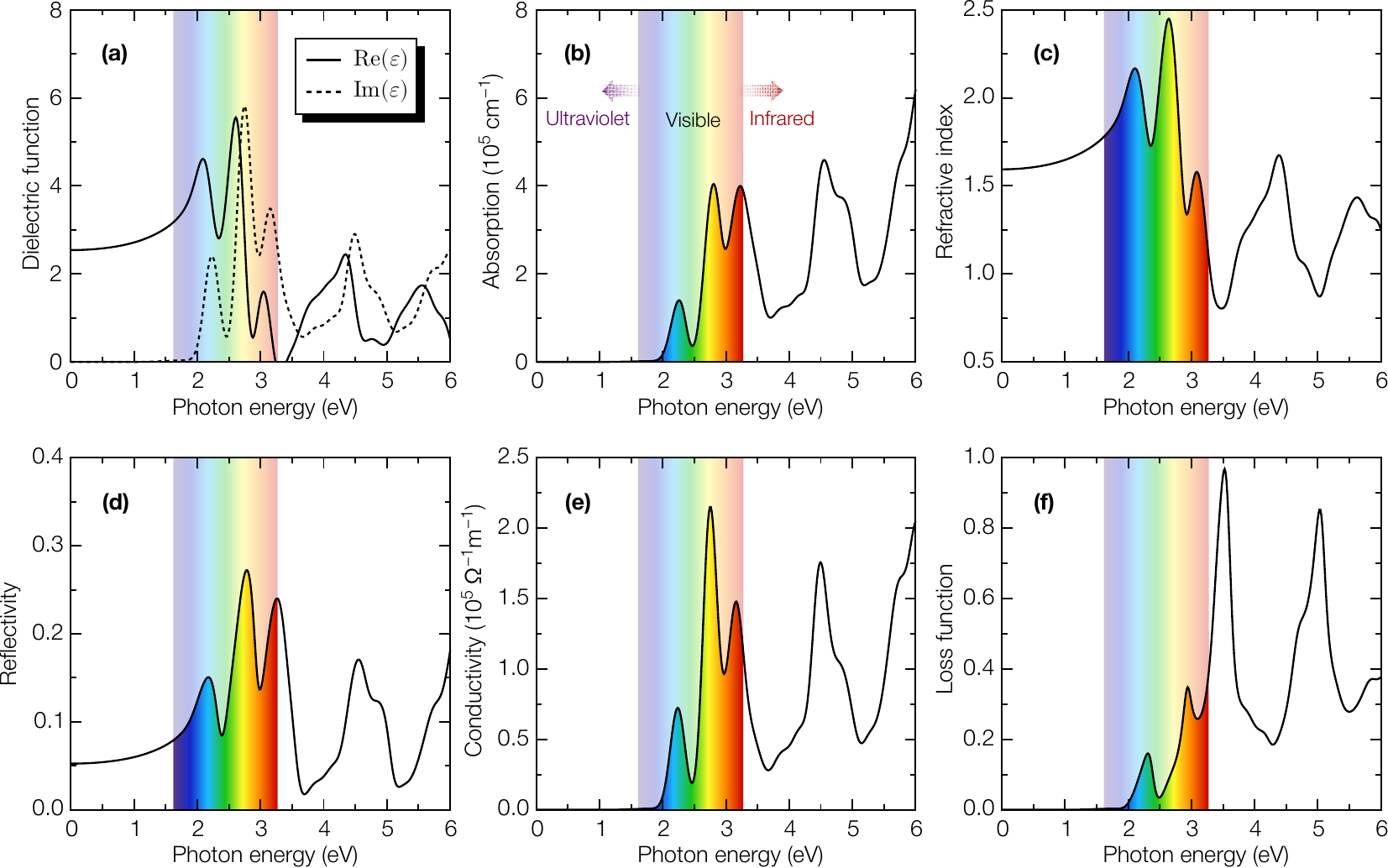}
\caption{Optical properties of the 4-5-6-8 nanoribbon. (a) Real and imaginary parts of the dielectric function. (b) Absorption coefficient spanning the near-visible and ultraviolet spectral regions. (c) Refractive index. (d) Reflectivity. (e) Optical conductivity. (f) Energy loss function. The shaded region denotes the visible range, emphasizing the spectral window relevant for optoelectronic applications.}
\label{fig:fig5}
\end{center}
\end{figure*}

Extending the multiphysical framework established above, the optical response displayed in Fig.~\ref{fig:fig5} reveals how the symmetry-broken bonding network governs light-matter interaction in the 4-5-6-8 nanoribbon. Optical excitations directly probe the joint density of states and the dipole-allowed interband transitions, thereby providing an integrated perspective on whether the nonequivalent polygons merely perturb the electronic band structure or reorganize it into a qualitatively distinct electronic medium. The latter clearly emerges as the physically consistent interpretation.

The imaginary component of the dielectric function, shown in Fig.~\ref{fig:fig5}(a), indicates that the onset of strong optical absorption occurs near the electronic gap, defining an optical gap in the visible range. This behavior contrasts with pristine graphene, whose gapless spectrum suppresses interband absorption at low photon energies, and instead aligns the present ribbon with semiconducting graphene nanoribbons where quantum confinement activates discrete optical channels. In quasi-one-dimensional carbon systems, strongly bound excitons are known to dominate the absorption spectrum due to reduced dielectric screening and enhanced Coulomb interactions, often shifting optical transitions relative to single-particle gaps \cite{prezzi2008optical}. Although the present results are obtained within an independent-particle framework, the spectral shape and sharpness of the first peaks are consistent with this confinement-driven optical landscape, reinforcing the interpretation that topology-induced band formation sets the stage for excitonic activity.

The absorption spectrum in Fig.~\ref{fig:fig5}(b) further clarifies this point. Multiple pronounced peaks populate the visible region and extend into the ultraviolet, demonstrating that the redistribution of $\pi$-orbital overlap across nonequivalent rings generates a hierarchy of optically active transitions rather than a single threshold. Such behavior mirrors the sensitivity of graphene nanoribbon optical transitions to width, edge configuration, and family classification, parameters known to control the emergence of discrete excitonic resonances and tunable optoelectronic response \cite{prezzi2008optical}. Here, however, the governing variable is not simply geometric confinement but the deliberate breaking of hexagonal symmetry. Topology, therefore, assumes a role analogous to width in conventional armchair ribbons, acting as a primary control knob for spectral engineering.

The refractive index and reflectivity, presented in Fig.~\ref{fig:fig5}(c) and (d), respectively, reinforce the presence of strong interband coupling within the visible range. Peaks in the refractive index approaching values above 2 indicate substantial electronic polarizability, whereas the moderate reflectivity suggests that incident radiation is preferentially absorbed rather than reflected. This combination is particularly desirable for nanoscale photonic and optoelectronic architectures, as it implies efficient electromagnetic coupling without excessive optical losses at interfaces.

Additional insight is obtained from the optical conductivity in Fig.~\ref{fig:fig5}(e), whose maxima coincide with the dominant absorption features. Because optical conductivity measures the ease with which photoexcited carriers contribute to current, these peaks reveal that the same topology-enabled states responsible for band formation also support efficient photocarrier generation. In this sense, the optical response is not an isolated property but a direct continuation of the transport landscape discussed earlier, confirming that electronic delocalization extends into the excited-state regime.

Finally, the loss function in Fig.~\ref{fig:fig5}(f) identifies collective electronic oscillations at higher photon energies, signaling the emergence of plasmon-like excitations typical of low-dimensional carbon systems. Such features originate from coherent charge-density fluctuations and are highly sensitive to the underlying band topology. Their presence suggests that the 4-5-6-8 framework may enable tunable plasmonic behavior, a possibility rarely accessible in conventional nanoribbons without external patterning or doping.

Taken as a unified optical portrait, these results demonstrate that breaking hexagonal symmetry does not merely open the electronic band gap; it restructures the entire excitation spectrum, spanning single-particle transitions, photocarrier generation, and collective modes. Topology thus operates as a cross-scale design principle linking atomic bonding to macroscopic optical observables. When viewed alongside the mechanical robustness, strain-tunable electronic structure, and predictable transport response established earlier, the optical behavior confirms that the 4-5-6-8 nanoribbon constitutes a genuinely multiproperty platform in which geometry is transformed into a governing physical parameter rather than a passive structural descriptor.

\section*{Conclusions}

We have demonstrated that the 4-5-6-8 carbon nanoribbon transcends the conventional interpretation of nonhexagonal rings as structural irregularities, establishing instead a materials architecture in which topology functions as a predictive design parameter. Periodic symmetry breaking reorganizes the bonding network into a stable, electronically coherent lattice that simultaneously exhibits mechanical robustness, a tunable electronic band structure, suppressed phonon transport, and strong optical activity.

Unlike standard graphene nanoribbons, where multifunctionality often relies on edge engineering, chemical modification, or extrinsic disorder, the present system embeds these capabilities directly within its geometric framework. Strain is converted into a controlled electronic response, thermal transport is intrinsically modulated by bond hierarchy, and optical transitions emerge from the redistributed orbital landscape. The ability of a single tight-binding parametrization to reproduce hybrid-functional trends further indicates that the governing physics responses are strongly determined by the topological.

These findings advance a broader principle for carbon-based nanomaterials: deliberately breaking hexagonal symmetry can transform geometry into a unifying physical variable that couples electronic, mechanical, thermal, and optical phenomena. The experimentally accessible 4-5-6-8 nanoribbon therefore represents more than a new member of the graphene family; it provides a blueprint for topology-driven multiproperty engineering and a pathway toward rationally designed low-dimensional systems in which functionality emerges as a direct consequence of the lattice architecture.

\section*{Methods}

First-principles calculations were carried out within the framework of density functional theory (DFT) using the Spanish Initiative for Electronic Simulations with Thousands of Atoms (SIESTA) code~\cite{soler2002,hohenberg1964,kohn1965,artacho1999}. Structural relaxations and initial electronic analyses employed the Perdew-Burke-Ernzerhof (PBE) generalized gradient approximation~\cite{perdew1996,perdew1997}, while the hybrid HSE06 functional~\cite{heyd2003hybrid,Heyd2006} was adopted to address the well-known electronic band gap underestimation of semilocal functionals. Hybrid-functional calculations were performed using the HONPAS package~\cite{Qin2015,Shang2020}, which preserves the SIESTA numerical framework.

Core electrons were described by norm-conserving Troullier-Martins pseudopotentials in the Kleinman-Bylander form~\cite{troullier1991,hamann1979,kleinman1982}. A double-$\zeta$ polarized basis set was used together with a kinetic energy cutoff of 700~Ry. Brillouin-zone sampling employed a Monkhorst-Pack grid of $20\times1\times1$~\cite{monkhorst1976}. Atomic positions and lattice vectors were fully relaxed until residual forces were below $0.001$~eV/\AA\ and the total energy converged to $10^{-5}$~eV. Periodic boundary conditions were applied along the ribbon axis, while vacuum regions of 35~\AA\ and 50~\AA\ were introduced along the transverse directions to suppress spurious interactions between periodic images.

Dynamical structural stability was evaluated through phonon dispersion calculations using finite displacements in sufficiently large supercells to eliminate artificial coupling. The absence of imaginary frequencies confirms that the symmetry-broken lattice corresponds to a true minimum of the potential energy surface. High-frequency modes around $\sim3200$~cm$^{-1}$ originate from C-H stretching vibrations, consistent with hydrogen-passivated carbon nanostructures.

Thermal structural stability was further assessed via \textit{ab initio} molecular dynamics  (AIMD) within the canonical ensemble using a Nosé-Hoover thermostat~\cite{evans1985nose}. Simulations performed at elevated temperatures preserved the structural integrity of the ribbon, indicating that the nonbenzenoid topology remains robust under thermal fluctuations.

To access length scales beyond the reach of first-principles methods, large-scale classical molecular dynamics simulations were conducted with the LAMMPS package~\cite{Thompson2022}. Atomic interactions were modeled using the second-generation reactive empirical bond-order (REBO) potential~\cite{Brenner2002}, extensively validated for carbon-based nanostructures \cite{felix2024irida,zhan2014structure,mortazavi2014multiscale,felix2025lattice,wei2022unusual}. The equations of motion were integrated using the velocity-Verlet algorithm~\cite{verlet1967computer}, considering a time step $dt = 0.1$ fs, for a total time of 5.0 nanoseconds, and kinetic energy swaps were performed every 2000 time steps. System lengths reached up to $\sim200$~nm, enabling the accurate description of long-wavelength phonons and realistic mechanical responses.

Lattice thermal transport was computed using the reverse non-equilibrium molecular dynamics (RNEMD) method proposed by Müller-Plathe~\cite{Muller1997}. Within this framework, the lattice thermal conductivity $\sigma_\text{ph}$ is obtained from Fourier’s law as: 

\begin{equation}
\sigma_\text{ph} (L_x) = - \frac{1}{\nabla_x T}
\left[
\frac{\sum_{\text{swaps}} \Delta K}{2 A \Delta t}
\right],
\end{equation}

where $\sum_{\text{swaps}} \Delta K$ is the total exchanged kinetic energy over the simulation time $\Delta t$, $\nabla_x T$ is the temperature gradient along the transport direction, and $A$ is the effective cross-sectional area of the system. The factor of 2 accounts for the bidirectional heat flux associated with periodic boundary conditions.

The length dependence of the thermal conductivity was analyzed through \cite{felix2018thermal}:

\begin{equation}
\frac{1}{\sigma_\text{ph}(L_x)} = \frac{1}{\sigma_\text{ph}^\infty} \left(1+\frac{\Lambda}{L_x}\right),
\label{eq:length-dependence}
\end{equation}

where $\sigma_\text{ph}^\infty$ is the intrinsic thermal conductivity in the diffusive limit and $\Lambda$ denotes the effective phonon mean free path.

The thermal conductance $\kappa_\text{ph} (L_x)$ was calculated from the length-dependent thermal conductivity $\sigma_\text{ph} (L_x)$ using the relation:

\begin{equation}
\kappa_\text{ph} (L_x) = \sigma_\text{ph}(L_x)\,\frac{A}{L_x},
\label{eq:conductance-conductivity}
\end{equation}

where $A$ is the effective cross-sectional area and $L_x$ is the system length along the transport direction. This definition allows a direct comparison between conductance and conductivity results and is particularly convenient for analyzing size effects in low-dimensional systems. The ballistic lattice thermal conductance $\kappa_{\mathrm{ph}}^{0}$ corresponds to the short-length limit of the length-dependent lattice thermal conductance $\kappa_{\mathrm{ph}}(L_x)$, namely $\kappa_{\mathrm{ph}}^{0} = \lim_{L_x \to 0} \kappa_{\mathrm{ph}}(L_x)$.

The optical response was obtained from the complex frequency-dependent dielectric function within the independent-particle approximation:

\begin{equation}
\varepsilon(\omega) = \varepsilon_1(\omega) + i\varepsilon_2(\omega).
\end{equation}

The imaginary part of the dielectric tensor was evaluated within the independent-particle approximation using the momentum-matrix-element formalism appropriate for periodic systems \cite{gajdovs2006linear}. It is expressed as

\begin{eqnarray}
\varepsilon_2^{\alpha\beta}(\omega)
&=&
\frac{4\pi^2 e^2}{\Omega}\nonumber\\
&&\times\sum_{v,c,\mathbf{k}}
w_{\mathbf{k}}
\,
\langle \psi_{c,\mathbf{k}} |
\hat{p}_{\alpha}
| \psi_{v,\mathbf{k}} \rangle
\langle \psi_{v,\mathbf{k}} |
\hat{p}_{\beta}
| \psi_{c,\mathbf{k}} \rangle
\nonumber \\
&&\times
\delta(E_{c,\mathbf{k}}-E_{v,\mathbf{k}}-\hbar\omega),
\end{eqnarray}

where $v$ and $c$ denote valence and conduction states, respectively, $w_{\mathbf{k}}$ is the weight of the $\mathbf{k}$ point, $\hat{p}_{\alpha}$ is the momentum operator along the Cartesian direction $\alpha$, and $\Omega$ represents the normalization volume of the simulation cell.

The real part was obtained through the Kramers-Kronig relation \cite{kronig1926,kramers1927}:

\begin{equation}
\varepsilon_1(\omega) =
1 +
\frac{2}{\pi}
\mathcal{P}
\int_0^\infty
\frac{\omega' \varepsilon_2(\omega')}
{\omega'^2-\omega^2}
\, d\omega',
\end{equation}

ensuring a causal optical response.

Because the nanoribbon is modeled within a three-dimensional periodic supercell with large vacuum regions, the computed dielectric response is artificially reduced by polarization dilution \cite{yadav2023first}. To recover the intrinsic quasi-one-dimensional behavior, the dielectric tensor was renormalized according to:

\begin{equation}
\varepsilon^{1D}(\omega)
=
1 +
\frac{L_y L_z}{A_{\mathrm{eff}}}
\left[
\varepsilon^{3D}(\omega)-1
\right],
\end{equation}

where $L_y$ and $L_z$ are the supercell dimensions perpendicular to the ribbon axis and $A_{\mathrm{eff}} = w d_{\mathrm{eff}}$ defines the effective cross-sectional area. Here, $w$ is the physical ribbon width, and an effective thickness of $d_{\mathrm{eff}} = 3.35$~\AA \cite{ni2007graphene}, corresponding to the graphite interlayer spacing, was adopted. This normalization removes the spurious dependence on vacuum size and enables quantitative comparison with other carbon-based systems.

The complex refractive index $\tilde{n}(\omega)=n(\omega)+ik(\omega)$ was obtained from the dielectric function as:

\begin{eqnarray}
&n(\omega)&=
\sqrt{
\frac{|\varepsilon(\omega)|+\varepsilon_1(\omega)}{2}
},
\qquad \nonumber \\
&k(\omega)&=
\sqrt{
\frac{|\varepsilon(\omega)|-\varepsilon_1(\omega)}{2}
},
\end{eqnarray}

from which the absorption coefficient follows as:

\begin{equation}
\alpha(\omega)=\frac{2\omega k(\omega)}{c},
\end{equation}

and the reflectivity at normal incidence as:

\begin{equation}
R(\omega)=
\left|
\frac{n(\omega)-1+ik(\omega)}
{n(\omega)+1+ik(\omega)}
\right|^2.
\end{equation}

The optical conductivity was evaluated through:

\begin{equation}
\sigma(\omega)=
-i\omega\varepsilon_0
\left[
\varepsilon(\omega)-1
\right],
\end{equation}

while the energy-loss function:

\begin{equation}
L(\omega)=
\Im
\left[
-\frac{1}{\varepsilon(\omega)}
\right]
=
\frac{\varepsilon_2(\omega)}
{\varepsilon_1^2(\omega)+\varepsilon_2^2(\omega)},
\end{equation}

was used to identify collective electronic excitations.

All spectra were computed considering light polarized along the longitudinal direction of the ribbon axis, thus capturing the intrinsic optical anisotropy of the quasi-one-dimensional lattice. A Gaussian broadening of 0.15~eV was introduced to mimic finite excited-state lifetimes and to smooth discrete interband transitions near the band edges.

Internal mixed geometries presented within the 4-5-6-8 nanoribbon promotes interesting quantum confinement. The quasi-one-dimensional architecture leads to the predominant formation of single-quantum transport channels. The corresponding nanoscopic properties of such nanostructured systems can be described using a single orbital Tight-Binding Hamiltonian:

\begin{equation}
H = \sum_{i} \epsilon \ c_i^\dagger c_i + \sum_{i,j} t_{ij} ( c_i^\dagger c_j + \text{H.c.} ),
\end{equation}

where $\epsilon_i$ represents the on-site energy,  $c_i^\dagger$ ($c_i$) is the creation (annihilation) operator for electrons at site $i$, and $t_{ij}$ is the hopping integral between the sites $i$ and $j$, which is calculated employing an exponential decaying as:

\begin{equation}
    t_{ij}=t_1 \ e^{-\beta  (\frac{r_{ij}}{d_{\text{min}}}-1 )},
\end{equation}

with $t_1$ being the hopping related to the first nearest-neighbor distance $r_{ij}$ between $i,j$ lattice sites, $d_{\text{min}}=\text{1.39}$ \AA \ is the nearest-neighbor distance, and $\beta$ is a fitting parameter that controls the hopping energy range. As the ratio $r_{ij}/d_{\text{min}}$ is always larger than one beyond the first neighborhood, small values of $\beta$ increase the number of neighbors with non-negligible hopping contributions in the description. 
The energy parameters are adjusted accordingly to DFT outputs; on-site term is chosen  $\epsilon=-0.9$ eV, the hopping energy $t_1=-3.85$ eV, and $\beta=2.8$. This parametrization scheme has been successfully employed for carbon-based mixed geometries in graphene allotropes, such as multidimensional biphenylene systems \cite{lage2024, LageSliding} and organic chains \cite{lage2025emergent}.    

Quantum transport properties are investigated using the Green's function $\mathcal{G}(E)$ within the Landauer formalism \cite{datta1997electronic} defined as:

\begin{equation}
\mathcal{G}(E) = \left[ E \mathbb{I} - H - \Sigma_L(E) - \Sigma_R(E) \right]^{-1},
\end{equation}

where $\Sigma_L(E)$ and $\Sigma_R(E)$ are the self-energies of the left and right electrodes, calculated at the energy $E$, respectively. The total density of states (DOS) is obtained by $\text{DOS}=-(1/\pi)\text{Im} \{ Tr [ \mathcal{G}(E) ] \}$. The transmission function $T(E)$ is given by\cite{datta1997electronic}:

\begin{equation}
\mathcal{T}(E) = \text{Tr} \left[ \Gamma_L(E) \mathcal{G}(E) \Gamma_R(E) \mathcal{G}^\dagger(E) \right],
\end{equation}

where $\Gamma_L(E)$ and $\Gamma_R(E)$ are the broadening matrices of the left and right electrodes, respectively. In this work, the leads are taken to be structurally identical to the scattering region as illustrated in Fig.~\ref{fig:fig1}(a).
The electronic current across the scattering region reads:

\begin{eqnarray}
    I&=&\frac{2e}{h} \int dE \ \mathcal{T}(E) \times \nonumber \\ && [ f_L(E,\mu,T)-f_R(E,\mu,T) ],
\end{eqnarray}

with $f_{L,R}(E,\mu,T)=[1+e^{(E-\mu_{L,R})/k_{B}T}]^{-1}$ being the Fermi-Dirac distributions of left and right leads, respectively. An applied bias, V, shifts the
left and right chemical potentials as $ \mu_{L,R} = E_F \pm eV/2$, with $E_F$ being the electrode Fermi energy. To pursue the spatial distribution of electronic transmission, we employ a local transmission coefficient $T^{\alpha}_{ij}(E)$ which computes the transmission between $j$ and $i$ sites, and leads $\alpha=L,R$ for a given energy \cite{lima2022local}, namely:

\begin{equation}
    T^{\alpha}_{ij}(E)=2\text{Im}\{ [ \mathcal{G}(E)\Gamma_\alpha \mathcal{G}^\dagger(E) ]_{ji}H_{ij}\}.
\end{equation}

Once we compute the electronic transmission $\mathcal{T}(E)$ for $T=0 $ K, the electronic conductance $G(\mu,T)$, Seebeck coefficient $S(\mu,T)$, power factor $PF(\mu,T)$, and electronic thermal conductance $\kappa_e(\mu,T)$ can be obtained in terms of standard generalized transport coefficient $\mathcal{L}_{n}(\mu,T)$, defined as \cite{grosso2014,rodrigues2022exploring}:

\begin{eqnarray}
\mathcal{L}_{n}(\mu,T)&=&-\frac{2}{h} \int dE \ \mathcal{T}(E) \times \nonumber \\&&  (E-\mu)^{n} \partial_E f(E,\mu,T).
\end{eqnarray}

The conductance $G(\mu,T)$, the Seebeck  $S(\mu,T)$ , and the other coefficients $PF(\mu,T)$,   and $  \kappa_e(\mu,T)$ can be derived as follows:

\begin{gather}
    G(\mu,T)=e^{2}\mathcal{L}_{0}(\mu,T),\\
    S(\mu,T)=-\frac{1}{eT}\frac{\mathcal{L}_{1}(\mu,T)}{\mathcal{L}_{0}(\mu,T)},\\
    PF(\mu,T)=S(\mu,T)^{2}\sigma(\mu,T),\\
    \kappa_e(\mu,T)=\Bigg(\mathcal{L}_{2}(\mu,T)-\frac{\mathcal{L}_{1}(\mu,T)^{2}}{\mathcal{L}_{0}(\mu,T)} \Bigg).
\end{gather}

Finally, the thermoelectric efficiency is determined by the figure of merit, expressed as:

\begin{equation}
    ZT=\frac{G(\mu,T) S(\mu,T)^2}{\kappa_{e}(\mu,T)+\kappa_{ph}(T)},
\end{equation}

where the $\kappa_{ph}$ is the phononic contribution to the thermoelectric conductance, which depends only with the temperature $T$. 

We use the defined quantities to analyze thermal and electronic properties, and the stability of 4-5-6-8 nanoribbons.

\begin{acknowledgement}
This work received partial support from the Brazilian Coordination for the Improvement of Higher Education Personnel (CAPES), the National Council for Scientific and Technological Development (CNPq), the Research Support Foundation of the Federal District (FAPDF), the São Paulo Research Foundation (FAPESP), and the National Institute of Organic Electronics (INEO). D.G.S. acknowledges financial support from CAPES under grant No. 88887.102348/2025-00 (Finance Code 001). D.S.G. acknowledges support from INEO, CNPq, and FAPESP under Grant No. 2025/27044-5. A.L. and L.L.L. acknowledge financial support from the National Institute of Science and Technology Nanocarbon and 2D Materials (INCT Nanocarbono e Materiais 2D). M.L.P.J. acknowledges financial support from FAPDF (grant 00193-00001807/2023-16), CNPq (grants 444921/2024-9 and 308222/2025-3), and CAPES (grant 88887.005164/2024-00).
\end{acknowledgement}


\bibliography{references}

\end{document}